\documentclass[letterpaper, 10 pt, conference]{ieeeconf}  

\IEEEoverridecommandlockouts                              

\usepackage{cite}
\usepackage{amsmath,amssymb,amsfonts}
\usepackage{bm}
\usepackage{graphicx}
\usepackage{algorithm}
\usepackage{algpseudocode} 
\usepackage{hyperref}
\usepackage{textcomp}
\usepackage{xcolor}
\usepackage{soul}
\let\labelindent\relax
\usepackage{enumitem}

\usepackage{verbatim}
\usepackage{amsthm}
\usepackage[english]{babel}
\newtheorem{theorem}{Theorem}
\newtheorem{guideline}{Guideline}
\newtheorem{remark}{Remark}
\DeclareMathOperator{\EX}{\mathbb{E}}

\makeatletter
\g@addto@macro\normalsize{%
  \setlength\abovedisplayskip{4.8pt}
  \setlength\belowdisplayskip{4.8pt}
  \setlength\abovedisplayshortskip{4.8pt}
  \setlength\belowdisplayshortskip{4.8pt}
  \setlength{\belowcaptionskip}{0pt}
\setlength{\abovecaptionskip}{0pt}
}
\makeatother

\title{\LARGE \bf
Design Guidelines for Nonlinear Kalman Filters via Covariance Compensation 
}

\author{Shida Jiang$^{1}$, Jaewoong Lee$^{1}$, Shengyu Tao$^{1}$, and Scott Moura$^{1}$
\thanks{*This work was not supported by any organization}
\thanks{$^{1}$Shida Jiang, Jaewoong Lee, Shengyu Tao, and Scott Moura are with the Department of Civil and Environmental Engineering,
        University of California, Berkeley, CA 94720, USA. Their emails are respectively
        {\tt\small shida\_jiang@berkeley.edu, ljw7696@berkeley.edu, shengyu.tao@chalmers.se, smoura@berkeley.edu}}%
}

\begin{document}

\maketitle
\thispagestyle{empty}
\pagestyle{empty}

\begin{abstract}

Nonlinear extensions of the Kalman filter (KF), such as the extended Kalman filter (EKF) and the unscented Kalman filter (UKF), are indispensable for state estimation in complex dynamical systems, yet the conditions for a nonlinear KF to provide robust and accurate estimations remain poorly understood. This work proposes a theoretical framework that identifies the causes of failure and success in certain nonlinear KFs and establishes guidelines for their improvement. Central to our framework is the concept of covariance compensation: the deviation between the covariance predicted by a nonlinear KF and that of the EKF. With this definition and detailed theoretical analysis, we derive three design guidelines for nonlinear KFs: (i) invariance under orthogonal transformations, (ii) sufficient covariance compensation beyond the EKF baseline, and (iii) selection of compensation magnitude that favors underconfidence. Both theoretical analysis and empirical validation confirm that adherence to these principles significantly improves estimation accuracy, whereas fixed parameter choices commonly adopted in the literature are often suboptimal. The codes and the proofs for all the theorems in this paper are available at \href{https://github.com/Shida-Jiang/Guidelines-for-Nonlinear-Kalman-Filters}{https://github.com/Shida-Jiang/Guidelines-for-Nonlinear-Kalman-Filters}.

\end{abstract}

\section{INTRODUCTION}

\subsection{Background and Literature Review}

The Kalman filter (KF) is the cornerstone of state estimation for linear dynamical systems. In its classical form, the KF provides recursive updates of both the state mean and covariance under linear dynamics with random noise. For additive noise that follows an arbitrary distribution, the KF is an optimal linear estimator that minimizes the sum of the mean-squared error of the state estimation \cite{book3}. The success of KF stems from this mathematical optimality, as well as its broad applicability in fields ranging from navigation to signal processing \cite{use1, use4}. However, the optimality of the KF relies heavily on the assumption of linearity, which rarely holds in real-world systems.

To extend the KF framework to nonlinear systems, researchers have developed several widely used variants. The extended Kalman filter (EKF) linearizes the nonlinear dynamics locally in each iteration, making it a simple and computationally efficient approach \cite{book1}. However, it is often inaccurate when nonlinearities are strong. The second-order EKF (EKF2) improves on this idea by incorporating higher-order terms in the approximation \cite{book2}. In parallel, the unscented Kalman filter (UKF) was introduced as an alternative that avoids explicit linearization. Instead, UKF selects sigma points that capture the first two moments of the state distribution and then uses the transformed sigma points to approximate the mean and covariance of states after the nonlinear transformation \cite{julier}. Note that UKF can either refer to a family of nonlinear KFs or a specific type of nonlinear KF. In this work, we adopt the former definition to avoid confusion. Different versions of the UKF have emerged, such as the spherical simplex Kalman filter (SKF) and the cubature Kalman filter (CKF), which differ mainly in how the sigma points are selected \cite{UKFreview}. The UKF generally achieves higher accuracy than the EKF, albeit with a slight increase in complexity. However, UKF has several hyperparameters and numerous variants, and an improper selection of them can negatively impact the algorithm's performance \cite{UKFreview}.

Despite the various types of algorithms, nonlinear state estimation remains a fundamentally challenging task. Given only the first two moments of the original states, the mean and covariance of the states after a nonlinear transformation cannot, in general, be uniquely determined. Even in the seemingly simple case of quadratic functions, this limitation persists for covariance estimation. For instance, if a random variable $X$ has mean zero and variance one, then $X^2$ always has mean one, but its variance can take any nonnegative value depending on the distribution of $X$. Consequently, filters that attempt to outperform the EKF must adopt certain assumptions about the underlying state distribution. Yet such assumptions are inherently fragile, as the shape of the distribution continually changes under nonlinear dynamics and measurements.

This inherent problem has hindered the development of nonlinear KF since the field's birth. On one hand, new algorithms continue to emerge that empirically outperform the EKF in many applications. On the other hand, it remains challenging to explain why certain nonlinear KFs succeed, while others fail. For example, it has been reported that the SKF often performs worse than the CKF, although they are both variants of UKF and share very similar ideas \cite{comparisonSKF, SSKF}. Similarly, a technique called `scaling', which adjusts the position of the sigma points, can substantially improve performance for some UKF variants (e.g., SKF) but has little effect for others (e.g., CKF) \cite{SSKF, comparison1, comparison2}. To date, no existing theory fully explains these puzzling empirical results. 

\subsection{Contributions}
This paper develops a new theoretical framework that explains why certain nonlinear KFs fail and provides guidelines for improving their performance. Since the EKF is the most fundamental nonlinear KF, we introduce the concept of \emph{covariance compensation}. Specifically, we define the covariance compensation matrix as the difference between the estimated covariance matrix produced by a given nonlinear KF and that produced by the EKF. We show that high-performing methods (e.g., CKF, EKF2) typically induce a positive semidefinite (PSD) compensation matrix, whereas fragile variants (e.g., SKF) may not. Building on this insight, we establish general guidelines for robust and accurate state and covariance estimation. In general, the main contributions of this paper are summarized as follows:
\begin{itemize}[leftmargin=*]
    \item We introduce the concept of covariance compensation and conduct a systematic analysis of the conditions required for accurate state and covariance estimation in nonlinear KFs. The proposed theory provides a solid explanation for the performance gap between seemingly similar algorithms (e.g., CKF vs. SKF).
    \item We derive principled guidelines for robust nonlinear KFs and validate them through both theoretical and empirical analysis.
    \item Our theoretical and empirical results call into question the long-standing Gaussianity assumption for system states—an assumption that has underpinned most nonlinear KF derivations for almost six decades.
\end{itemize}

\section{COVARIANCE COMPENSATION IN NONLINEAR KALMAN FILTERS}\label{section_review}
\subsection{General Framework for Nonlinear Kalman Filters}

Different types of nonlinear KFs differ only in how they approximate the first two moments of the states after a nonlinear transformation, while sharing the same underlying framework. Conventionally, this framework mirrors that of the linear KF and consists of two steps per iteration: ``predict'' and ``update.'' In the ``predict'' step, the first two moments of the states are propagated through the state transition functions. In the ``update'' step, these predictions are corrected using the measurement functions and the observed measurements. Although this framework has been in use for over sixty years, our previous work \cite{frame} has demonstrated that it systematically underestimates the actual state covariance matrix, resulting in degraded state estimation accuracy over time. While rigorous mathematical statements can be found in \cite{frame}, the intuition is that the conventional equations assume the Kalman gain is optimal. However, the Kalman gain, which was optimized based on the approximated measurement function near the predicted states, generally will not minimize the trace of the actual covariance.
As a result, the filter tends to overestimate the Kalman gain's effectiveness, leading to overconfident state estimates.

To mitigate this issue, our previous work proposed extending the framework with two additional steps: ``recalibrate'' and ``back out.'' The ``recalibrate'' step re-approximates the system around the updated states, allowing the filter to assess the actual effect of the selected Kalman gain on the covariance. Because the gain is derived from approximations, the trace of the covariance matrix may increase after the update and recalibration if the system is strongly nonlinear and the predicted states have large variances. In such cases, the update fails to improve estimation accuracy. To handle this situation, the ``back out'' step enables the filter to revert to its prior state when the update is unhelpful. The detailed equations and pseudo-code of the proposed framework are provided in Algorithm~\ref{new_algorithm}.

\begin{algorithm}
\footnotesize
	\caption{The framework for nonlinear Kalman filters used in this paper}\label{new_algorithm}
	\begin{algorithmic}[1]
        \Statex \textbf{Input:} Process noise covariance matrix $\boldsymbol{Q}_k$, Measurement noise covariance matrix $\boldsymbol{R}_k$, state transition function $\boldsymbol{f}(\boldsymbol{x},\boldsymbol{u})$, measurement function $\boldsymbol{h}{(\boldsymbol{x})}$, system inputs $\boldsymbol{u}_k$, measurements $\boldsymbol{z}_k$
        \Statex \textbf{Initialization:}
        \State $\boldsymbol{\hat{x}}_{0|0}=\EX[\boldsymbol{x}_0]$
        \State $\boldsymbol{P}_{0|0}=\EX[(\boldsymbol{\hat{x}}_{0|0}-\boldsymbol{x}_0)(\boldsymbol{\hat{x}}_{0|0}-\boldsymbol{x}_0)^T]$
		\For {every time step $k$}
        \Statex \hspace{1em} \textbf{Predict:}
        \State Estimate $\boldsymbol{\hat{x}}_{k|k-1}$ and $\boldsymbol{P}_{k|k-1}$
        \Statex \hspace{1em} (based on $\boldsymbol{\hat{x}}_{k-1|k-1}, \boldsymbol{P}_{k-1|k-1}, \boldsymbol{f}, \boldsymbol{u}_k,$ and $\boldsymbol{Q}_k$)
		\Statex \hspace{1em} \textbf{Update:}
        \State Estimate $\boldsymbol{\hat{z}}_{k|k-1}, \boldsymbol{P}_{xz,k|k-1}$ and $\boldsymbol{P}_{z,k|k-1}$ 
        \Statex \hspace{1em} (based on $\boldsymbol{\hat{x}}_{k|k-1}, \boldsymbol{P}_{k|k-1},$ and $\boldsymbol{h}$)
        \State $\boldsymbol{S}_{k|k-1}=\boldsymbol{P}_{z,k|k-1}+\boldsymbol{R}_k$
        \State $\boldsymbol{K}_k=\boldsymbol{P}_{xz,k|k-1}\boldsymbol{S}_{k|k-1}^{-1}$
        \State $\boldsymbol{\hat{x}}_{k|k}=\boldsymbol{\hat{x}}_{k|k-1}+\boldsymbol{K}_k(\boldsymbol{z}_k-\boldsymbol{\hat{z}}_{k|k-1})$
		\Statex \hspace{1em} \textbf{Recalibrate:}
        \State Estimate $\boldsymbol{P}_{xz,k|k}$ and $\boldsymbol{P}_{z,k|k}$
        \Statex \hspace{1em} (based on $\boldsymbol{\hat{x}}_{k|k}, \boldsymbol{P}_{k|k-1},$ and $ \boldsymbol{h}$)
        \State $\boldsymbol{S}_{k|k}=\boldsymbol{P}_{z,k|k}+\boldsymbol{R}_k$
        \State {$\boldsymbol{P}_{k|k}=\boldsymbol{P}_{k|k-1}+\boldsymbol{K}_k\boldsymbol{S}_{k|k}\boldsymbol{K}_k^T-\boldsymbol{P}_{xz,k|k}\boldsymbol{K}_k^T-\boldsymbol{K}_k\boldsymbol{P}_{xz,k|k}^T$}
        \Statex \hspace{1em} \textbf{Back out:}
        \If{$\text{tr}(\boldsymbol{P}_{k|k})>\text{tr}(\boldsymbol{P}_{k|k-1})$}
        \State $\boldsymbol{\hat{x}}_{k|k}=\boldsymbol{\hat{x}}_{k|k-1}$
        \State $\boldsymbol{P}_{k|k}=\boldsymbol{P}_{k|k-1}$
        \EndIf
		\EndFor
        \Statex where, $\boldsymbol{\hat{x}}$ is the estimated states, $\boldsymbol{P}$ is the states' covariance matrix, $\boldsymbol{\hat{z}}$ is the estimated measurements, $\boldsymbol{P}_{xz}$ is the covariance between the states and the measurements, $\boldsymbol{P}_{z}$ is the covariance of the estimated measurements, $\boldsymbol{S}$ is the innovation (or residual) covariance, $\boldsymbol{K}_k$ is the Kalman gain. The subscript $k|k-1$ represents the estimated value before the state update, and $k|k$ represents the estimated value after the state update. The framework degrades to the conventional one when $\boldsymbol{P}_{xz,k|k}=\boldsymbol{P}_{xz,k|k-1}$ and $\boldsymbol{P}_{z,k|k}=\boldsymbol{P}_{z,k|k-1}$.
	\end{algorithmic}
    \normalsize
    
\end{algorithm}
\vspace{-1em}

\subsection{The Moment Estimation Problem}\label{moment_sec}
In nonlinear KFs, both the state transition functions and the measurement functions can be nonlinear. Therefore, the problem of estimating the first two moments of the states after a nonlinear transformation is addressed three times (once in each of the first three steps) per iteration in the nonlinear KF framework we introduced earlier. The differences between various types of nonlinear Kalman filters are fully characterized by how they address the problem of estimating the first two moments. Without loss of generality, the problem can be formulated as follows. 

Consider random vectors $\bm{x}\in \mathbb{R}^n,\bm{z}\in \mathbb{R}^m$ and a nonlinear measurable function mapping $f:\mathbb{R}^n\rightarrow\mathbb{R}^m$ that satisfy:
\begin{equation}\label{setup2}
    \bm{x}\sim(0,I_{n\times n}),\bm{z}=f(\bm{x}),
\end{equation}
where $\bm{x}\sim(0, I_{n\times n})$ means that $\bm{x}$ has a mean of zero and a covariance of the identity matrix. Denote the mean of $\bm{z}$, the covariance of $\bm{z}$, and the cross covariance between $\bm{x}$ and $\bm{z}$ as $\bar{\bm{z}}$, $P_z$, and $P_{xz}$, respectively. The problem is to estimate $\bar{\bm{z}}$, $P_z$, and $P_{xz}$ given the conditions in (\ref{setup2}).

Note that the problem statement above is equivalent to the general case. Namely, if $\bm{x}$ has a mean of $\bm{\bar{x}}$ and a covariance of $P_x$, we can write $\bm{x}=\bm{\bar{x}}+L\bm{u}$, where $\bm{u}$ has zero mean and a covariance of the identity matrix, and  $LL^T=P_x$ is the Cholesky factorization of $P_x$. With this affine transformation, any functions of $\bm{x}$ can be considered as functions of $\bm{u}$. Therefore, in the latter analysis, it is sufficient for us to consider the simple case represented by (\ref{setup2}).

\subsection{Formal definition of covariance compensation}
EKF estimates $\bar{\bm{z}}$, $P_z$, and $P_{xz}$ by \cite{book2}:
\begin{equation}\label{EKF}
    \begin{cases}
    \bar{\bm{z}}^{\textnormal{EKF}}= f(\bm{0}) \\
    P_{z}^\textnormal{EKF}=f'(\bm{0})(f'(\bm{0}))^T \\
    P_{xz}^\textnormal{EKF}=(f'(\bm{0}))^T
    \end{cases},
\end{equation}
where $f'(\bm{0})$ is the $m$ by $n$ Jacobian matrix at $\bm{x}=\bm{0}$.

As we previously mentioned, the covariance compensation matrix is defined as the difference between the estimated covariance matrix given by a specific type of KF and the estimated covariance matrix given by the EKF. With (\ref{EKF}), we can define the covariance compensation matrix $P_{com}$ as:
\begin{equation}\label{Pcom}
    P_{com}=P_{z}^\textnormal{est}-P_{z}^\textnormal{EKF}=P_{z}^\textnormal{est}-f'(\bm{0})(f'(\bm{0}))^T,
\end{equation}
where $P_{z}^\textnormal{est}$ is the covariance estimation made by the algorithm of interest.

\subsection{Covariance Compensation in Second-Order EKFs}
Assuming (\ref{setup2}), EKF2 estimates $\bar{\bm{z}}$, $P_z$, and $P_{xz}$ by \cite{book2}:
\begin{equation}\label{EKF2}
    \begin{cases}
    \bar{\bm{z}}^{\textnormal{EKF2}}= f(\bm{0}) + \tfrac{1}{2}[\textnormal{tr}(f_i''(\bm{0}))]_i\\
    P_{z}^\textnormal{EKF2}=f'(\bm{0})(f'(\bm{0}))^T+\tfrac{1}{2}[\textnormal{tr}(f_i''(\bm{0})f_j''(\bm{0}))]_{ij} \\
    P_{xz}^\textnormal{EKF2}=(f'(\bm{0}))^T
    \end{cases},
\end{equation}
where $f_i''(\bm{0})$ is the $n$ by $n$ Hessian matrix of $f_i$ at $\bm{x}=\bm{0}$. From the definition, the covariance compensation matrix used in EKF2 is:
\begin{equation}\label{Gau}
    P_{com}^{\textnormal{EKF2, Gau}}=\tfrac{1}{2}[\textnormal{tr}(f_i''(\bm{0})f_j''(\bm{0}))]_{ij}\succeq 0.
\end{equation}
Note that the matrix is PSD because it is a Gram matrix under the Frobenius inner product. The superscript ``Gau'' represents the Gaussian distribution assumption, which is necessary to derive this equation. If $\bm{x}$ follows some other distributions, the covariance compensation matrix will also be different. For example, when $\bm{x}$ is evenly distributed on the sphere $||\bm{x}||_2^2=n$, it's not hard to verify that $\bm{x}$ has zero mean and a variance of the identity matrix, and the covariance compensation matrix in this case can be calculated as:
{\small
\begin{equation}\label{sphere}
    P_{com}^{\textnormal{EKF2, Sphere}}=\frac{n[\textnormal{tr}(f_i''(\bm{0})f_j''(\bm{0}))-\tfrac{1}{n}\textnormal{tr}(f_i''(\bm{0}))\textnormal{tr}(f_j''(\bm{0}))]_{ij}}{2(n+2)}\succeq 0.
\end{equation}}
The derivation of (\ref{sphere}) can be found in the \href{https://github.com/Shida-Jiang/Guidelines-for-Nonlinear-Kalman-Filters}{Supplementary Material on Github}. As we previously mentioned, in nonlinear KF, the shape of the distribution of the states changes when either the state-transition functions or the measurement functions are nonlinear. Therefore, there is no reason to stick to the Gaussian distribution assumption. In fact, our empirical results indicate that (\ref{sphere}) can be used as a substitute for (\ref{Gau}), and the state estimation result will be mostly similar. Therefore, we introduce an additional parameter $\beta$ to characterize the magnitude of the covariance compensation matrix for EKF2, and rewrite (\ref{Gau}) as:
\begin{equation}
    P_{com}^{\textnormal{EKF2}}(\beta)=\frac{\beta}{4}[\textnormal{tr}(f_i''(\bm{0})f_j''(\bm{0}))]_{ij}.
\end{equation}
Clearly, when $\beta=2$, $P_{com}^{\textnormal{EKF2}}(\beta)$ will be the same as $P_{com}^{\textnormal{EKF2,Gau}}$. Since $P_{com}^{\textnormal{EKF2,Gau}}$ is PSD, a larger $\beta$ means to perform more covariance compensation and make the estimated covariance larger.
\subsection{Covariance Compensation in Unscented Kalman Filters}
\begin{figure}[htbp]
\centering
\includegraphics[width=7.2cm]{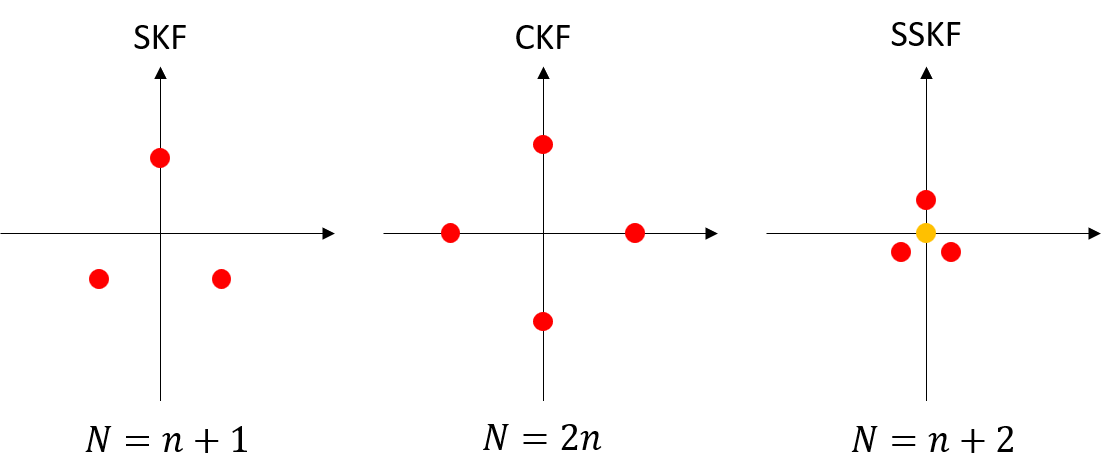}
\caption{The selected sigma points in different variants of unscented Kalman filters when $n=2, \bm{x}\sim(\bm{0},I_{2\times 2})$. }
\label{points}
\vspace{-1em}
\end{figure}

As pointed out by Menegaz et al. in \cite{UKFreview}, UKF has many variants and many different notations. In this paper, we primarily discuss the variants of UKF covered in Julier's most highly cited paper \cite{julier}, which include SKF, CKF, Scaled SKF (SSKF), and Scaled CKF (SCKF). Different variants of UKF mainly differ in the way of sampling the ``sigma points'', which characterize the discrete distribution that the states are assumed to follow. We denote the total number of sigma points as $N$. For SKF, CKF, SSKF, and SCKF, $N$ is respectively equal to $n+1$, $2n$, $n+2$, and $2n+1$. Note that SSKF has lower computational complexity than SCKF. Yet, they provide the same state and covariance estimations, differing only in their computation methods \cite{SSKF}. Therefore, it is sufficient for us to only discuss SKF, CKF, and SSKF in the remainder of this subsection. When $n=2, \bm{x}\sim(\bm{0},I_{2\times 2})$, the sigma points selected by different variants of UKF are shown in Fig. \ref{points}.

In a more general case, denote the set of sigma points as $\{\bm{\xi}_i\}_{i=1,2,\ldots,N}$. For different variants of UKF, the sampled sigma points $\bm{\xi}_i$ can be written as \cite{julier,SSKF}:
\begin{equation}
    \bm{\xi}^\textnormal{SKF}_i=\sqrt{n+1}\bm{\xi}_i'-\frac{(n+1+\sqrt{n+1})}{n(n+1)}\bm{1}_{n\times 1},i=1,\ldots ,n+1,
\end{equation}
\begin{equation}
    \bm{\xi}^\textnormal{CKF}_i=
    \begin{cases}
    \sqrt{n}\bm{\xi}_i',&i=1,\ldots ,n\\
    -\sqrt{n}\bm{\xi}_{i-n}',&i=n+1,\ldots ,2n
    \end{cases},
\end{equation}
\begin{equation}\label{SSKF_points}
    \bm{\xi}^\textnormal{SSKF}_i=
    \begin{cases}
    \alpha\bm{\xi}^\textnormal{SKF}_i,&i=1,\ldots ,n+1\\
    \bm{0},&i=n+2
    \end{cases},
\end{equation}
where $\alpha$ is a tiny number (e.g., 0.001), $\bm{1}_{n\times 1}$ is a vector of ones, and $\bm{\xi}_i'$ is defined as:
\begin{equation}
    [\bm{\xi}_1',\bm{\xi}_2',\ldots,\bm{\xi}_n']=I_{n\times n},\quad\bm{\xi}_{n+1}'=\frac{\bm{1}_{n\times 1}}{\sqrt{n+1}-1}.
\end{equation}
SKF and CKF both assume that the states follow a uniform discrete distribution. With this assumption, SKF and CKF estimates $\bar{\bm{z}}$, $P_z$, and $P_{xz}$ by:
\begin{equation}\label{SKF}
    \begin{cases}
    \bar{\bm{z}}^{\textnormal{SKF/CKF}}= \tfrac{1}{N}\sum_{i=1}^{N}f(\bm{\xi}^{\textnormal{SKF/CKF}}_i) \\
    P_{z}^\textnormal{SKF/CKF}=\tfrac{1}{N}\sum_{i=1}^{N}(f(\bm{\xi}^{\textnormal{SKF/CKF}}_i)- \bar{\bm{z}}^{\textnormal{SKF/CKF}})(\cdot)^T\\
    P_{xz}^\textnormal{SKF/CKF}=\tfrac{1}{N}\sum_{i=1}^{N}\bm{\xi}^{\textnormal{SKF/CKF}}_i(f(\bm{\xi}^{\textnormal{SKF/CKF}}_i)- \bar{\bm{z}}^{\textnormal{SKF/CKF}})^T
    \end{cases}
\end{equation}
where the notation $(\cdot)^T$ means the transpose of the preceding term. With these equations, the covariance compensation matrices for SKF and CKF can be calculated as:
\begin{equation}
    P_{com}^{\textnormal{SKF/CKF}}=P_{z}^\textnormal{SKF/CKF}-f'(\bm{0})(f'(\bm{0}))^T.
\end{equation}
Note that the equations for SKF and CKF are almost the same, and they only differ in the value of $N$ and the set of sigma points they sample. However, as suggested by the following theorem, CKF has a PSD covariance compensation matrix when the nonlinear function is real analytic at the mean of the states. In contrast, SKF doesn't necessarily have this property. 
\begin{theorem}\label{theorem1}
With the conditions in (\ref{setup2}), and when the nonlinear mapping $f$ is real analytic at the origin, we have $P_{com}^{\textnormal{CKF}}\succeq 0$. (Proof in the Supplementary Material on Github.)
\end{theorem}
On the other hand, SSKF assumes that all the sigma points except the center point have the same weight, as indicated by the color in Figure \ref{points}. Given the location of the sigma points in (\ref{SSKF_points}) and the constraint that the covariance is the identity matrix, we can calculate that the weights of the points are:
\begin{equation}\label{SKFweight}
    w_i^{\textnormal{SKF}}(\alpha)=
    \begin{cases}
        \frac{1}{(n+1)\alpha^2},&i=1,2,\ldots,n+1\\
        \frac{\alpha^2-1}{2\alpha^2},&i=n+2
    \end{cases}.
\end{equation}
With (\ref{SKFweight}), SSKF estimates $\bar{\bm{z}}$, $P_z$, and $P_{xz}$ by:
\begin{equation}\label{SSKF}
    \begin{cases}
    \bar{\bm{z}}^{\textnormal{SSKF}}= \sum_{i=1}^{n+2}w_if(\bm{\xi}^{\textnormal{SSKF}}_i) \\
    P_{z}^\textnormal{SSKF}=P_{z,0}^{\textnormal{SSKF}}+\sum_{i=1}^{n+2}w_i(f(\bm{\xi}^{\textnormal{SSKF}}_i)- \bar{\bm{z}}^{\textnormal{SSKF}})(\cdot)^T\\
    P_{xz}^\textnormal{SSKF}=\sum_{i=1}^{n+2}w_i\bm{\xi}^{\textnormal{SSKF}}_i(f(\bm{\xi}^{\textnormal{SSKF}}_i)- \bar{\bm{z}}^{\textnormal{SSKF}})^T
    \end{cases},
\end{equation}
where 
\begin{equation}
    P_{z,0}^{\textnormal{SSKF}}=(1-\alpha^2+\beta)(f(\bm{0})- \bar{\bm{z}}^{\textnormal{SSKF}})(f(\bm{0})- \bar{\bm{z}}^{\textnormal{SSKF}})^T.
\end{equation}
From (\ref{SSKF}), we can see that the estimated $\bar{\bm{z}}$ and $P_{xz}$ are directly calculated from the assumed discrete distribution. However, the estimated $P_{z}$ has two parts: the covariance calculated from the assumed distribution and an additional PSD term, $P_{z,0}$. The primary reason for introducing this term is to guarantee that $P_{z}$ is PSD. Namely, in (\ref{SKFweight}), the weight of the final point is negative when $\alpha$ is a small number. Such a negative weight can make the calculated covariance not PSD without additional compensation. 

While the equations used in SSKF appear complicated, existing literature \cite{relationship, SSKF} proved that when $\alpha \rightarrow 0$, (\ref{SSKF}) will be equivalent to:
\begin{equation}\label{limit}
    \begin{cases}
    \bar{\bm{z}}^{\textnormal{SSKF}}= f(\bm{0}) + \tfrac{1}{2}[\textnormal{tr}(f_i''(\bm{0}))]_i\\
    P_{z}^\textnormal{SSKF}=f'(\bm{0})(f'(\bm{0}))^T+\frac{\beta}{4}[\textnormal{tr}(f_i''(\bm{0}))\textnormal{tr}(f_j''(\bm{0}))]_{ij} \\
    P_{xz}^\textnormal{SSKF}=(f'(\bm{0}))^T
    \end{cases}, 
\end{equation}
which is very similar to the equations for EKF2 in (\ref{EKF2}). Therefore, a common practice is also to select $\beta=2$ to represent the Gaussian distribution \cite{UKF_b}. With (\ref{limit}), the covariance compensation matrix in SSKF can be written as:
\begin{equation}\label{beta_SSKF}
    P_{com}^{\textnormal{SSKF}}(\beta)=\frac{\beta}{4}[\textnormal{tr}(f_i''(\bm{0}))\textnormal{tr}(f_j''(\bm{0}))]_{ij}.
\end{equation}

Finally, we want to extend the equations for SKF and CKF in (\ref{SKF}) so that their covariance compensation matrix also contains $\beta$, allowing us to adjust the magnitude. Comparing (\ref{SSKF}) and (\ref{SKF}), we can see that SKF is a special case of SSKF when $\beta=0,\alpha=1$. If we fix $\alpha=1$ and adjust $\beta$, we can add additional covariance compensation to SKF without scaling the sigma points. In this way, we can rewrite the estimated $P_{z}$ in SKF and CKF as:
\begin{equation}\label{SKF_beta}
    P_{z}^\textnormal{SKF*/CKF*}(\beta)=P_{z}^\textnormal{SKF/CKF}+\beta(f(\bm{0})- \bar{\bm{z}}^{\textnormal{SKF/CKF}})(\cdot)^T.
\end{equation}
Note that we use the ``*'' sign to represent the existence of additional covariance compensation ($\beta> 0$) in SKF and CKF. Especially, according to \cite{relationship}, when $f$ represents quadratic functions, (\ref{SKF_beta}) can also be written as:
\begin{equation}\label{SKF_beta2}
    P_{z}^\textnormal{SKF*/CKF*}(\beta)=P_{z}^\textnormal{SKF/CKF}+\frac{\beta}{4}[\textnormal{tr}(f_i''(\bm{0}))\textnormal{tr}(f_j''(\bm{0}))]_{ij}.
\end{equation}

\section{GUIDELINES FOR ROBUST AND ACCURATE STATE ESTIMATION}\label{method}
\subsection{Applications of Covariance Compensation}\label{method1}

As we summarized in the previous section, except for the EKF, different nonlinear KFs often rely on different assumptions about the state distribution. Before comparing these assumed distributions, we first analyze what distributions are indistinguishable to the KF algorithm by revisiting the problem formulation introduced in Section~\ref{moment_sec}. There, we mentioned that the problem setup in~(\ref{setup2}) is general, since we can always perform an affine transformation to standardize a random vector to have zero mean and unit covariance. However, this standardization is not unique. Specifically, let $\bm{x}$ satisfy $E[\bm{x}]=\bm{0}$ and $\mathrm{Cov}(\bm{x})=I_n$, and consider an affine transformation
\begin{equation}
    g(\bm{x}) = A\bm{x}+\bm{b}.
\end{equation}
Then $g(\bm{x})$ has the same first two moments as $\bm{x}$ if and only if $\bm{b}=\bm{0}$ and $AA^T=I_n$, i.e., $A$ is orthogonal. Therefore, the normalization method is not unique, as left-multiplying by an orthogonal matrix preserves zero mean and unit covariance. Meanwhile, the vector normalization technique was widely used in the derivation of various nonlinear KFs (e.g., EKF2 and UKF \cite{relationship}). Since the normalization is not unique, a natural question is: Do different normalizations affect the mean and covariance estimates?

\begin{figure}[htbp]
\centering
\includegraphics[width=7.5cm]{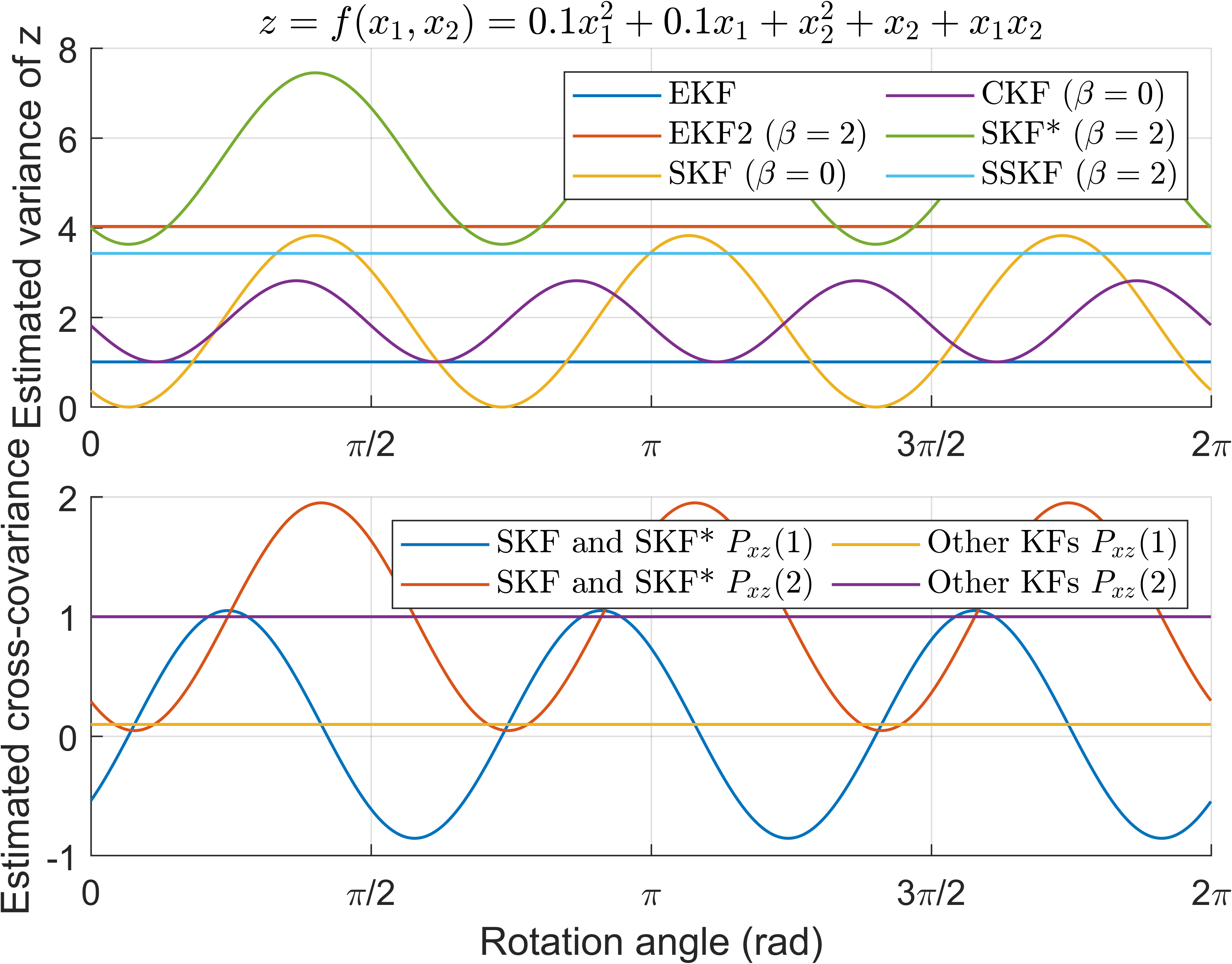}
\caption{The sensitivity of the estimated covariance $P_{z}^\textnormal{est}$ and $P_{xz}^\textnormal{est}=[P_{xz}(1) \quad P_{xz}(2)]^T$ to rotation in different nonlinear KFs. }
\label{problem_SKF}
\vspace{-2em}
\end{figure}

Ideally, the answer should be negative, i.e. the filter estimates are invariant w.r.t. normalization. In Fig. \ref{points}, for example, this invariance means that the estimated $\bar{\bm{z}}$, $P_{z}$, and $P_{xz}$ (denoted as $\bar{\bm{z}}^{\textnormal{est}}$, $P_{z}^\textnormal{est}$, and $P_{xz}^\textnormal{est}$) should be insensitive to the rotation of the sigma points. However, this is generally not the case for SKF and CKF, as shown in Fig. \ref{problem_SKF}.

Fig.~\ref{problem_SKF} illustrates how a rotation of the assumed state distribution (i.e., left-multiplying different orthogonal matrices) influences the covariance estimation $P_{z}^\textnormal{est}$ and $P_{xz}^\textnormal{est}$. For the EKF, the update equation is independent of the state distribution, so the covariance estimate remains invariant to the rotation angle. For the EKF2, the assumed standard Gaussian distribution is itself rotationally invariant, leading again to invariance. For the SSKF, although its sigma points are affected by rotation, the resulting estimates remain unchanged because the sigma points are infinitesimally close to the origin and have zero third- and higher-order moments. In contrast, both the CKF and SKF produce estimates that are sensitive to rotation.  

Intuitively, the high sensitivity of SKF to orthogonal transformations can substantially impair its state estimation accuracy. As shown in Fig.~\ref{problem_SKF}, SKF (yellow) may sometimes yield $P_z^{\textnormal{est}}$ values that are nearly zero, making the estimator severely overconfident. Such overconfidence is particularly problematic because an underestimated state covariance matrix prevents the filter from incorporating new measurements effectively in later iterations. In other words, even if the SKF underestimates $P_z$ in one iteration and overestimates it in the next, the net effect still tends to make the estimator overconfident and less accurate. While the CKF also exhibits this issue, its covariance is strictly lower-bounded by $P_{z}^\textnormal{EKF}$ according to the PSD property in Theorem~\ref{theorem1}. Since the EKF captures the dominant first-order nonlinearity, this lower bound usually provides reasonable accuracy, thereby making the sensitivity problem less severe for CKF.  

These observations motivate the following guideline for robust state estimation. According to our earlier analysis, this guideline implies that the covariance estimate should not be too small after an orthogonal transformation of the state. Although the guideline is motivated by intuition, a mathematically more rigorous justification will be provided in Remark \ref{remark2} in the following subsection.
\begin{guideline}\label{g1}
    For any transformation $f$, the covariance compensation matrix $P_{com}$, as defined in \eqref{Pcom}, should satisfy
    \begin{equation}
        P_{com}\succeq 0.
    \end{equation}
\end{guideline}

When the states follow some specific types of distributions, it is possible to derive a tighter bound for $P_{com}$. Specifically, if the states follow a radially symmetric distribution, we have the following theorem:
\begin{theorem}\label{theoremnew}
Consider the conditions in (\ref{setup2}), and further assume that $\bm{x}$ follows a radially symmetric distribution, i.e., its probability density depends only on $\|\bm{x}\|^2_2$.  
Let $f:\mathbb{R}^n\to\mathbb{R}^m$ be a vector of quadratic functions of the form
\begin{equation}\label{strong_assum}
    \bm{z}=f(\bm{x})=\bm{c}+f'(\bm{0})\bm{x}+\tfrac{1}{2}[\bm{x}^T f_i''(\bm{0}) \bm{x}]_i.
\end{equation}
Then the following equations hold:
\begin{equation}\label{thm_ekf}
        \begin{cases}
            \bar{\bm{z}}=f(\bm{0}) + \tfrac{1}{2}\,[\textnormal{tr}(f_i''(\bm{0}))]_i \\
            P_{z}\succeq f'(\bm{0})(f'(\bm{0}))^T+P_{com}^{\textnormal{EKF2, Sphere}} \\
            P_{xz}=(f'(\bm{0}))^T
        \end{cases},
\end{equation}
where $P_{com}^{\textnormal{EKF2, Sphere}}$ is defined in (\ref{sphere}). (Proof in the Supplementary Material on Github.)
\end{theorem}
We then show that this specific case can be extended to the general case via an orthogonal transformation. Namely, consider the following randomized orthogonal transformation:
\begin{equation}\label{ortho}
    g(\bm{x};\hat{A})=\hat{A}\bm{x},
\end{equation}
where $\hat{A}\in\mathbb{R}^{n\times n}$ is a random orthogonal matrix satisfying $\hat{A}\hat{A}^{T}=I$ and is independent of $\bm{x}$. We further take $\hat{A}$ to be Haar-uniform on the orthogonal group $O(n)$ (i.e., ``uniformly random'' over all orthogonal matrices). Then the transformed random vector $\tilde{\bm{x}}:=g(\bm{x};\hat{A})$ is rotationally invariant: its distribution is invariant under any deterministic orthogonal rotation, and its density (if exists) depends only on $\|\tilde{\bm{x}}\|_{2}$.

In two dimensions, for example, $\hat{A}$ corresponds to a rotation by a random angle uniformly distributed on $[0,2\pi)$. For the distributions in Fig.~\ref{points}, \eqref{ortho} rotates the sigma points, which in this case trace out a circle. More generally, any distribution with zero mean and unit covariance can be converted to a radially symmetric one using $g(\bm{x};\hat{A})$. Therefore, given Theorem~\ref{theoremnew}, we provide the following guideline:

\begin{guideline}\label{g2}
    If $f$ is quadratic and the states are normalized randomly using \eqref{ortho}, the estimates of $\bar{\bm{z}}$, $P_{xz}$, and $P_{z}$ given by the algorithm, after averaging over the Haar-random $\hat{A}$, should satisfy \eqref{thm_ekf}.
\end{guideline}

\begin{remark}\label{remark:g2_interpret}
Theorem~\ref{theoremnew} provides a systematic explanation for a well-known behavior of the EKF. Namely, when the mapping has a nontrivial quadratic component, the covariance estimates given by the EKF do not satisfy the inequality in (\ref{thm_ekf}), so the EKF often underestimates the state covariance.

Guideline~\ref{g2} shows that, after random normalization \eqref{ortho}, a robust sigma-point rule should (on average over $\hat{A}$) reproduce the EKF2 mean and cross-covariance for quadratic functions, and yield covariance compensation no smaller than the spherical benchmark \(P_{com}^{\textnormal{EKF2, Sphere}}\).
In particular, both CKF and SKF with \(\beta=0\) place their sigma points on the sphere \(\|\bm{x}^{(i)}\|_2^2=n\). Therefore, under \eqref{ortho}, each rotated point $\hat{A}\bm{x}^{(i)}$ is uniformly distributed on \(\{\bm{x}:\|\bm{x}\|_2^2=n\}\). Consequently, their rotation-averaged estimates coincide with the spherical EKF2 benchmark for quadratic maps.

Finally, \eqref{ortho} is best viewed as a diagnostic randomization for robustness checking rather than a default preprocessing step.
In practice, using a fixed normalization and fixed sigma points (as in CKF, SKF, and SSKF) avoids additional Monte-Carlo variability in $\hat{A}$ and is typically cheaper to implement.
\end{remark}

As shown in Fig.~\ref{problem_SKF}, both increasing covariance compensation and applying scaling can improve the performance of the SKF. Specifically, increasing $\beta$ raises the lower bound of $P_z^{\textnormal{est}}$, thereby alleviating overconfidence, while scaling reduces the sensitivity of SKF to orthogonal transformations, improving covariance estimation in the worst case. To better illustrate the power of covariance compensation, we present a simple example of a linear KF. This example shows how randomness in $P_z$ affects estimation accuracy, and how increasing $\beta$ can mitigate this effect. Consider the following one-dimensional system with no inputs:
\begin{equation}\label{setup_simple}
    x_k = x_{k-1} + w_{k-1}, \quad z_k = x_k + v_k,
\end{equation}
where $w_k$ is zero-mean Gaussian process noise with variance $10^{-8}$, and $v_k$ is zero-mean Gaussian measurement noise with variance $10^{-4}$. The initial state estimate is unbiased with variance $1$, and the measurement Jacobian equals $1$. 

\begin{figure}[t]
\centering
\includegraphics[width=7.5cm]{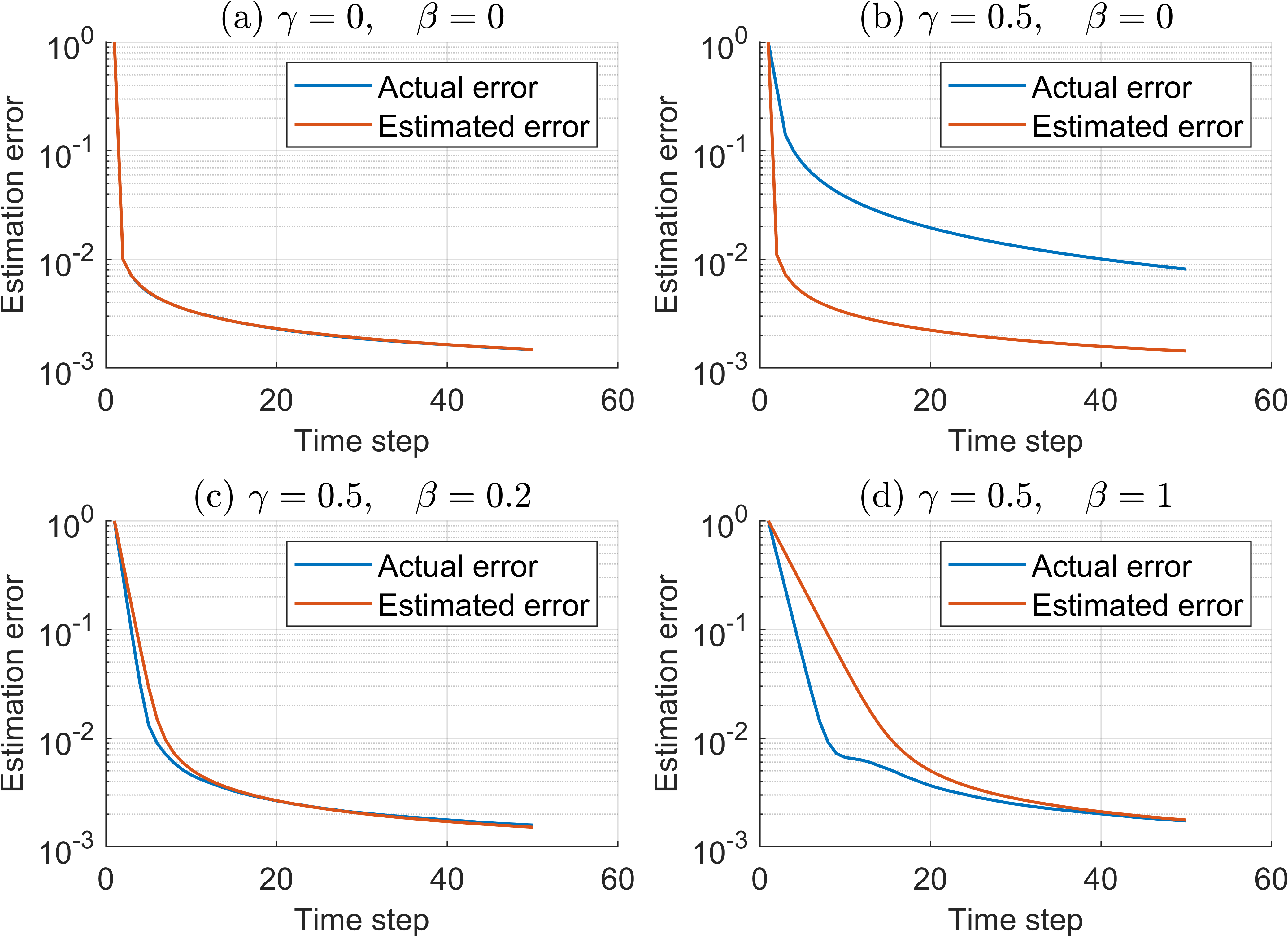}
\caption{Estimation error of a simple system under different magnitudes of covariance estimation fluctuations and covariance compensation.}
\label{why_compensation}
\vspace{-2em}
\end{figure}

Suppose, however, that the KF does not use the exact Jacobian, but instead updates the state using a noisy version. Let $\hat x_{k|k-1}$ and $P_{k|k-1}$ denote the predicted state and variance, respectively. The update is given by
\begin{equation}
    \hat x_{k|k} = \hat x_{k|k-1} + \frac{P_{k|k-1}\hat H}{(1+\beta)\hat H P_{k|k-1}\hat H + 10^{-4}}\,(z_k - \hat x_{k|k-1}),
\end{equation}
where $z_k$ is the measurement, $\beta$ is the covariance compensation magnitude, and $\hat H$ is a random variable representing the inaccurate Jacobian, uniformly distributed on $[1-\gamma,\,1+\gamma]$. The state estimation results under different setups are shown in Fig.~\ref{why_compensation}. In Fig.~\ref{why_compensation}, the \emph{actual error} denotes the RMSE averaged over 10{,}000 simulations, while the \emph{estimated error} denotes $\sqrt{P_{k|k}}$. In the ideal case (Fig.~\ref{why_compensation}(a)), these two errors coincide, meaning that the estimated variance is accurate. However, when zero-mean Jacobian errors are introduced (Fig.~\ref{why_compensation}(b)), the KF becomes significantly less accurate and overconfident. This situation mirrors the behavior of SKF, whose covariance estimates are highly sensitive to orthogonal transformations. Increasing the covariance compensation magnitude alleviates this issue (Fig.~\ref{why_compensation}(c)), but an excessively large $\beta$ can degrade performance (Fig.~\ref{why_compensation}(d)). These observations motivate an optimization algorithm for selecting the optimal $\beta$, introduced in the following subsection.

\subsection{Optimization of Covariance Compensation}\label{theory}
In this subsection, we model the selection of $\beta$ as an optimization problem. That is, we aim to find the optimal $\beta$ that minimizes the trace of the actual state covariance matrix after the state update. To formulate this optimization problem, we first need to derive the equation for the actual state covariance matrix after the state update.

Consider the following nonlinear dynamic system:
\begin{equation}\label{setup}
    \bm{x}_k=f(\bm{x}_{k-1},\bm{u}_{k-1})+\bm{w}_{k-1}, \quad
    \bm{z}_k=h(\bm{x}_k)+\bm{v}_k,
\end{equation}
where $\bm{x}_k$ are the actual states at the time step $k$, $\bm{u}_k$ are the inputs at the time step $k$, $\bm{w}_k$ are the process noises that follow a zero-mean multivariate distribution with covariance $Q_k$, and $\bm{v}_k$ are the measurement noises that follow a zero-mean multivariate distribution with covariance $R_k$. At an arbitrary time step $k$, the predicted states are denoted as $\hat{\bm{x}}_{k|k-1}$, and the errors $\hat{\bm{x}}_{k|k-1}-\bm{x}_{k}$ are assumed to follow a zero-mean multivariate distribution with covariance $P_{k|k-1}$. Then, in the ``update'' step, the predicted measurements are denoted as $\hat{\bm{z}}_{k|k-1}$, whose errors $\hat{\bm{z}}_{k|k-1}-h(\bm{x}_k)$ are assumed to follow a zero-mean multivariate distribution with covariance $P_{z,k}$. The actual cross-covariance matrix between the predicted states and the predicted measurements is denoted as $P_{xz,k}$. The differences between the predicted and actual measurements are called the measurement residuals, denoted as $\tilde{\bm{z}}_k$. Namely:
\begin{equation}\label{resi}
    \tilde{\bm{z}}_k:=\bm{z}_k-\hat{\bm{z}}_{k|k-1}.
\end{equation}
Since both $\bm{v}_k$ and $h(x_k)-\hat{\bm{z}}_{k|k-1}$ are zero-mean, the innovation $\tilde{z}_k$ is also zero-mean. If, in addition, the measurement noise $\bm{v}_k$ is assumed to be independent of the (noise-free) prediction error $h(\bm{x}_k)-\hat{\bm{z}}_{k|k-1}$, the covariance of the innovation is
\begin{equation} \label{residual_cov}
    S_{k}=P_{z,k}+R_k.
\end{equation}
The KF update for the state estimate is
\begin{equation}\label{state_update}
    \hat{\bm{x}}_{k|k}=\hat{\bm{x}}_{k|k-1}+K\Tilde{\bm{z}}_k,
\end{equation}
where $\hat{\bm{x}}_{k|k}$ are the updated states, and $K$ is the Kalman gain. With (\ref{resi}--\ref{state_update}), the covariance of $\hat{\bm{x}}_{k|k}$ is computed as \cite{derivation1}:
\begin{equation}\label{state_variance_true}
    P_{k|k}=P_{k|k-1}+KS_{k}K^T-P_{xz,k}K^T-KP_{xz,k}^T.
\end{equation}
KFs select the Kalman gain to minimize the trace of $P_{k|k}$. Specifically, the trace is minimized when:
\begin{equation}\label{Kalmangain}
    K_{op}=P_{xz,k}S_k^{-1},
\end{equation}
where $K_{op}$ is the theoretical optimal value of the Kalman gain. However, (\ref{Kalmangain})  cannot be directly applied to systems with nonlinear measurement functions. The primary reason is that $P_{z,k}, S_k$, and $P_{xz,k}$ can only be calculated if the distribution of the states is known. Since nonlinear KF only tracks the first two moments of the states, the best we can do is to approximate their values using a nonlinear KF variant. We denote these approximated covariance matrices as $\textcolor{red}{P_{z,k|k-1}}, \textcolor{red}{S_{k|k-1}}$, and $\textcolor{red}{P_{xz,k|k-1}}$ since they are all estimated based on the prediction. In Section \ref{section_review}, we have shown how different nonlinear KFs computes $\textcolor{red}{P_{z,k|k-1}}$ and $\textcolor{red}{P_{xz,k|k-1}}$. For example, the equations used in EKF2 are given in (\ref{EKF2}). After $\textcolor{red}{P_{z,k|k-1}}$ is computed, $\textcolor{red}{S_{k|k-1}}$ can be simply calculated:
\begin{equation}
    \textcolor{red}{S_{k|k-1}}=\textcolor{red}{P_{z,k|k-1}}+R_k.
\end{equation}

Before proceeding, it's necessary to make some additional assumptions to facilitate the theoretical analysis of these matrices. Namely, we interpret the approximated covariance matrices
$\textcolor{red}{P_{z,k|k-1}}$, $\textcolor{red}{S_{k|k-1}}$, and
$\textcolor{red}{P_{xz,k|k-1}}$ produced by a nonlinear KF as
(random) estimators of the actual quantities $P_{z,k}$, $S_k$, and $P_{xz,k}$.
More precisely, we regard $\textcolor{red}{P_{xz,k|k-1}}$ as a random matrix,
$\textcolor{red}{P_{z,k|k-1}}$ as a random PSD matrix, and
$\textcolor{red}{S_{k|k-1}}$ as a random positive definite matrix. For clarity, all such random covariance matrices are marked in red in this section. Additionally, we assume that these approximated covariance matrices are unbiased estimators and satisfy
\begin{equation}\label{expcondition}
    \begin{cases} \EX[\textcolor{red}{P_{z,k|k-1}}(\beta_0)] = P_{z,k} \\ \EX[\textcolor{red}{S_{k|k-1}}(\beta_0)] =S_k \\ \EX[\textcolor{red}{P_{xz,k|k-1}}]= P_{xz,k}\end{cases} .
\end{equation}
Nonlinear KFs select the Kalman gain as:
\begin{equation}\label{Kalmangain_real}
    \textcolor{red}{K_{est}}(\beta)=\textcolor{red}{P_{xz,k|k-1}}(\textcolor{red}{S_{k|k-1}}(\beta))^{-1}.
\end{equation}
The actual value of the updated state covariance matrix $\textcolor{red}{P_{k|k,ac}}$ given this selected Kalman gain can be calculated by substituting (\ref{Kalmangain_real}) into (\ref{state_variance_true}). Namely:
\begin{equation}\label{P_true}
\begin{aligned}
    &\textcolor{red}{P_{k|k,ac}}(\beta)=P_{k|k-1}+\textcolor{red}{P_{xz,k|k-1}}\textcolor{red}{S_{k|k-1}^{-1}}S_{k}\textcolor{red}{S_{k|k-1}^{-1}}\textcolor{red}{P_{xz,k|k-1}^T}\\
    &\quad -P_{xz,k}\textcolor{red}{S_{k|k-1}^{-1}}\textcolor{red}{P_{xz,k|k-1}^T}-\textcolor{red}{P_{xz,k|k-1}}\textcolor{red}{S_{k|k-1}^{-1}}P_{xz,k}^T.
\end{aligned}
\end{equation}
By definition, the trace of $\textcolor{red}{P_{k|k,ac}}$ is the expectation of the sum of the squared errors of the state estimates. Therefore, we can formulate the optimization problem as:
\begin{equation}\label{object}
    \min_\beta\mathbb{E}[\textnormal{tr}(\textcolor{red}{P_{k|k,ac}})].
\end{equation}
To simplify (\ref{P_true}), define $\textcolor{red}{P}:=\textcolor{red}{P_{xz,k|k-1}}S_k^{-\tfrac{1}{2}}$, $\textcolor{red}{S}:=S_k^{-\tfrac{1}{2}}\textcolor{red}{S_{k|k-1}}S_k^{-\tfrac{1}{2}}$. According to (\ref{expcondition}) and (\ref{P_true}),
\begin{equation}\label{expcondition2}
\EX[\textcolor{red}{P}] = \bar{P}= P_{xz,k}S_k^{-\tfrac{1}{2}}, \quad
    \EX[\textcolor{red}{S}(\beta_0)] = I. 
\end{equation}
\begin{equation}
\begin{aligned}
    &\textcolor{red}{P_{k|k,ac}}(\beta)=P_{k|k-1}+\textcolor{red}{P}(\textcolor{red}{S}(\beta))^{-2}\textcolor{red}{P^T}\\
    &\quad -\bar{P}(\textcolor{red}{S}(\beta))^{-1}\textcolor{red}{P^T}-\textcolor{red}{P}(\textcolor{red}{S}(\beta))^{-1}\bar{P}^T.
\end{aligned}
\end{equation}
From the previous section, we can see that $\textcolor{red}{S}(\beta)$ is a linear function of $\beta$ and has the following form:
\begin{equation}\label{sb}
    \textcolor{red}{S}(\beta)=\textcolor{red}{S}(\beta_0)+(\beta-\beta_0)\textcolor{red}{\Delta S}.
\end{equation}

Since the actual values of $P_{xz,k}$ and $S_k$ are unknown, an explicit closed-form solution of $\beta$ for the optimization problem (\ref{object})–(\ref{sb}) cannot be derived. Nevertheless, two noteworthy special cases arise when 
\begin{equation}
    \textbf{Case 1: }\textcolor{red}{S}(\beta_0) \equiv I, 
    \quad 
    \Delta S = I.
\end{equation}
\begin{equation}
    \textbf{Case 2: }\textcolor{red}{P} \equiv \bar P\neq 0, 
    \quad 
    \textcolor{red}{\Delta S} = \textcolor{red}{\textcolor{red}{S}}(\beta_0)\succ 0.
\end{equation}
These two scenarios both correspond to the following question: if the best estimate of $S_k$ is attained by selecting $\beta = \beta_0$, does $\beta_0$ also minimize (\ref{object})?  Perhaps unexpectedly, the answer is negative, as established in the following theorem.
\begin{theorem}\label{theorem2}
(i) In case 1, the global minimizer $\beta^*$ of (\ref{object}) exists (finite if $\bar P\neq 0$, and infinite otherwise) and satisfies $\beta^*\ge \beta_0$. Additionally, when $\|\bar P\|_F^2\geq 0$, $\min f$ and $\beta^*$ both increase monotonically as $\mathbb{E}[\|\textcolor{red}{\Delta P}\|_F^2]$ increases. (ii) In case 2, the global minimizer $\beta^*$ of (\ref{object}) exists (possibly infinite) and satisfies $\beta^*\geq \beta_0$. (Proof in the Supplementary Material on Github.)
\end{theorem}

\begin{remark}\label{remark2}
    Theorem \ref{theorem2} leads to two important conclusions.  
    First, $\min f$ increases as $\mathbb{E}[\|\textcolor{red}{\Delta P}\|_F^2]$ increases, indicating that a high sensitivity of $P_{xz}^\textnormal{est}$ to orthogonal transformations can degrade state estimation performance. This underscores the importance of scaling, which mitigates such sensitivity.  
Second, the optimal $\beta$ does not necessarily yield the most accurate $\textcolor{red}{S_{k|k-1}}(\beta)$, but instead tends to overestimate the actual covariance. This finding justifies the need for a PSD covariance compensation matrix when the original covariance estimation is unbiased (i.e., $\beta_0=0$). Besides, it also provides a practical rule of thumb for selecting $\beta$ in nonlinear KFs and motivates the following guideline:
\end{remark}

\begin{guideline}\label{g3}
    The covariance compensation magnitude $\beta$ should be chosen such that the KF slightly overestimates the actual state covariance. In practice, this condition can be verified using the Normalized Innovation Squared metric (or similar metrics assessing the consistency of the innovation covariance estimation).
\end{guideline}

\section{EXPERIMENTAL VALIDATION}
To validate the effectiveness of the proposed guidelines, we consider three representative applications of nonlinear KFs: 3D target tracking, terrain-referenced navigation, and synchronous generator state estimation. A brief overview of these applications is provided in Table \ref{setup_brief}, while detailed system models and parameter settings can be found in \cite{frame}.

\begin{table}[htbp]\scriptsize
\vspace{-1em}
\centering
\caption{Nonlinear systems investigated in this paper.}\label{setup_brief}
\begin{tabular}{|c|c|c|c|}
\hline
Applications & \begin{tabular}[c]{@{}c@{}}Tracking\end{tabular} & Navigation & Generator \\ \hline
Number of states & 6 & 2 & 4 \\ \hline
Linear state transition function? & Yes & Yes & No \\ \hline
Number of measurements & 2 & 1 & 1 \\ \hline
Linear measurement functions? & No & No & No \\ \hline
Number of inputs & 3 & 2 & 3 \\ \hline
Number of iterations & 30 & 100 & 100 \\ \hline
Measurement std. & 0.01 m & $1$ m & 0.0001 \\ \hline
\end{tabular}
\vspace{-1em}
\end{table}

\begin{figure}[htbp]
\centering
\includegraphics[width=7cm]{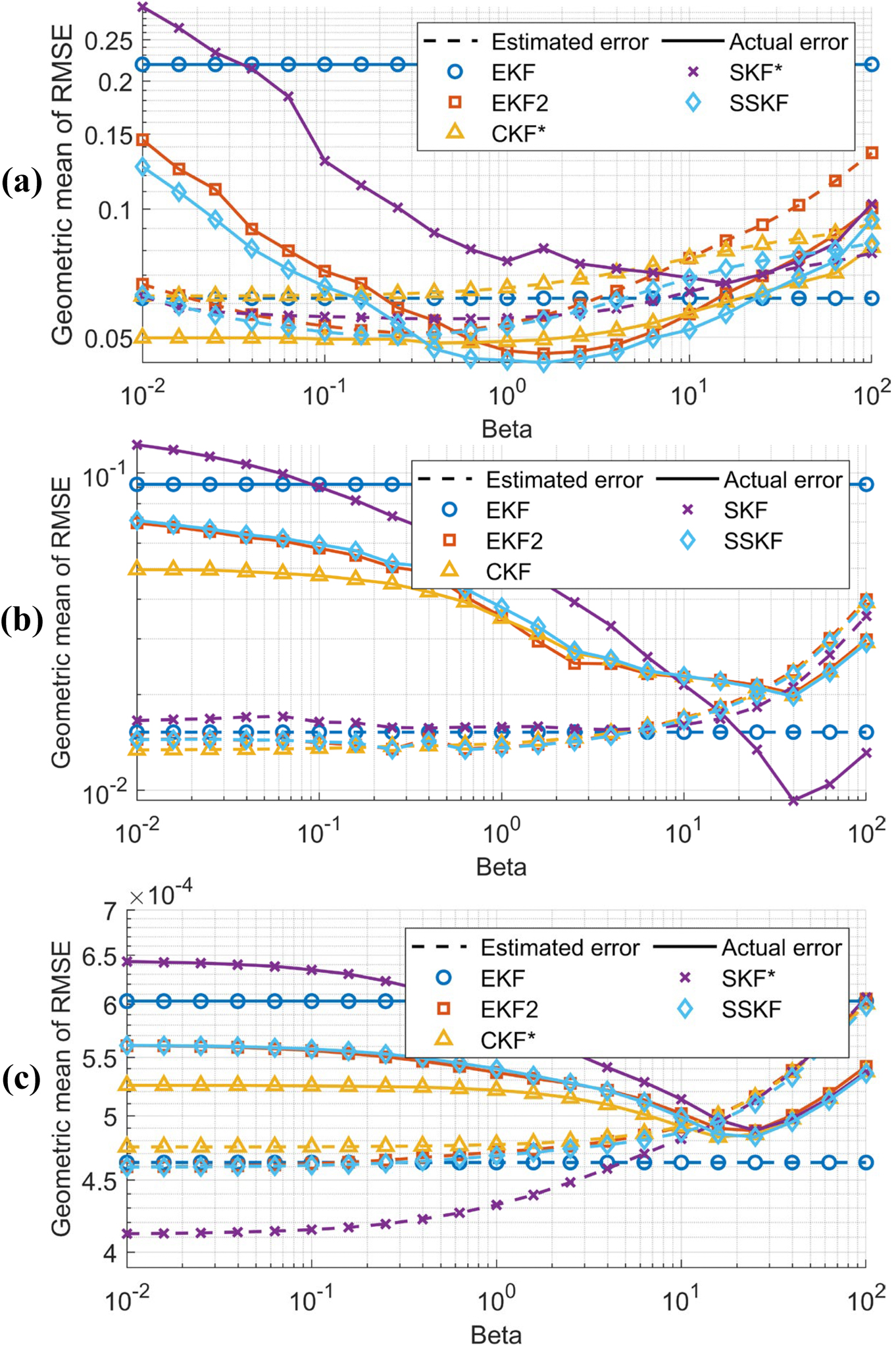}
\caption{The actual and estimated geometric mean of the state estimation RMSE under different magnitudes of covariance compensation. (a) 3D target tracking, (b) terrain-referenced navigation, (c) generator state estimation.}
\label{exp}
\vspace{-1em}
\end{figure}

The performance of state estimation using different nonlinear KFs is illustrated in Fig. \ref{exp}. Performance is assessed by first computing the state estimation RMSE at different timesteps, and then taking the geometric mean of the RMSE across all states and timesteps to characterize the filter's overall accuracy. The geometric mean, rather than the arithmetic mean, is used to eliminate discrepancies caused by differing units across states. In the figure, different nonlinear KFs are distinguished by colors and markers, while dotted and solid lines denote the estimated and actual errors, respectively. Here, the actual error refers to the deviation between the estimated and true state, whereas the estimated error corresponds to the square root of the estimated variance. To ensure fairness and reduce the effect of randomness, each KF is simulated 10,000 times for each value of $\beta$. Accordingly, both the estimated and actual RMSE are obtained from these 10,000 simulations, and they should be almost identical if the covariance estimation is accurate.

As shown in Fig. \ref{exp}, increasing $\beta$ monotonically inflates the estimated covariance and thus the estimated RMSE. In contrast, consistent with Fig. \ref{why_compensation}, the actual RMSE decreases at first and then increases.
Comparing the estimated and actual errors, we observe that the actual RMSE is minimized when the estimator is slightly underconfident. This conclusion holds consistently across all three applications and all four advanced nonlinear KFs considered in this paper (EKF2, SKF*, CKF*, and SSKF), underscoring the importance of Guideline \ref{g3}. In contrast, setting $\beta=0$ for CKF and $\beta=2$ for others according to convention is generally suboptimal. For the examples shown in Fig. \ref{exp}, the optimal $\beta$ that minimizes the actual RMSE ranges from 0.4 to 40, depending on the application and filter type, indicating that no fixed choice of $\beta$ achieves minimum RMSE in all cases.  

When $\beta \rightarrow 0$, the performance of EKF2 and SSKF becomes nearly identical. In fact, as shown in (\ref{EKF2}) and (\ref{limit}), the two algorithms coincide in this limit when the scaling factor $\alpha \rightarrow 0$. The slight discrepancy in Fig.~\ref{exp}(a) arises because the implementation uses $\min \beta=0.01$, which is not sufficiently close to zero. Besides, across all three cases, CKF* consistently delivers the best performance when $\beta\rightarrow 0$, which reflects its compliance with Guideline \ref{g2} and thus highlights the importance of this condition. Conversely, SKF* is consistently the worst-performing algorithm in this regime, validating the necessity of Guideline \ref{g1}.

\section{CONCLUSIONS}

This paper proposed a new perspective on nonlinear Kalman filter design by introducing the concept of covariance compensation. We showed that certain properties related to the covariance compensation matrix distinguish reliable algorithms (e.g., CKF, EKF2) from fragile ones (e.g., EKF, SKF). We distilled this insight into three guidelines for designing nonlinear KFs: invariance under orthogonal transformations, sufficient compensation beyond the EKF2-sphere baseline, and selecting $\beta$ to overestimate uncertainty. Experiments across multiple applications confirm that these guidelines improve the filter's accuracy and robustness. Future work includes extending the framework to other types of filters (e.g., particle filter) and developing adaptive strategies to tune $\beta$ online. 








\bibliographystyle{IEEEtran}
\bibliography{ref}

\end{document}